# ETUDE DES SIGNAUX RECUEILLIS PAR UN RADAR EMBARQUE SUR UN VEHICULE EN DEPLACEMENT. APPLICATION A L'INTERPRETATION DES SIGNAUX RECOLTES PAR LE RADAR WISDOM DE LA MISSION SPATIALE EXOMARS

F. Demontoux*, G. Ruffié*, Ph. Paillou**, C. Caruncho*, Julien Lahoudere**
* Université Bordeaux 1 - Laboratoire IMS - UMR 5218- 16 av Pey-Berland 33607 Pessac
**Laboratoire L3AB UMR5804  2 rue de l'Observatoire, B.P.89, 33270 Floirac,

## 1   Introduction

Le programme Aurora [1] débuté en 2001 fait partie de la stratégie de l'Europe pour l'exploration du système solaire. Il a été approuvé par le Conseil de Recherche de l'Union Européenne et le Conseil de l'Agence Européenne de l'espace en 2001.  L'objectif est d'explorer le Système Solaire, de stimuler le développement de nouvelles technologies et d'inspirer aux jeunes d'Europe un plus grand intérêt pour la science et la technologie.

Un premier objectif d'Aurora est de créer, puis d'appliquer, un plan européen à long terme pour l'exploration robotique et humaine de Mars. Un deuxième objectif est de rechercher la vie au delà de la Terre : les futures missions dans le cadre du programme utiliseront des outils sophistiqués pour étudier la possibilité de l'existence de formes de vie sur d'autres mondes dans le Système Solaire.

La mission ExoMars du programme Aurora de l'ESA a ainsi pour objectif d'envoyer un véhicule de type rover sur Mars en 2011. Au-delà des aspects technologiques de la mission, le rover emportera également une charge utile scientifique pour l'analyse du sous-sol de Mars, à la recherche de traces de vie passée et/ou présente dans le sol martien.

Le rover embarquera ainsi une foreuse qui permettra l'accès à des échantillons du sous-sol de Mars jusqu'à une profondeur de 2 mètres. Ceci constitue la grande originalité d'ExoMars, car le sous-sol de Mars reste encore inconnu et semble le meilleur endroit pour abriter des conditions favorables à la vie. Cette foreuse sera guidée par un système radar sondeur UHF, l'instrument WISDOM, qui permettra également de sonder le sous-sol de Mars jusqu'à quelques mètres de profondeur pour détecter de l'eau et étudier les structures géologiques.

L'équipe de Planétologie du L3AB-Bordeaux est partie prenante dans la réalisation du radar sondeur WISDOM et dans l'analyse de ses données, et prend en charge les modélisations géo-électriques et électromagnétiques avec le laboratoire PIOM (caractérisation électromagnétique d'échantillons de roches et modélisation numérique) et le Lunar and Planetary Institute de Houston (hydrogéologie martienne). Des modèles géo-électriques des premiers mètres du sous-sol de Mars ont ainsi été réalisés. Ils vont permettre de spécifier les données techniques du radar et de développer des méthodes d'inversion des données.

Nous disposons de modèles électromagnétiques analytiques (type IEM, GOM, SPM) et numériques (type FDTD et HFSS) pour simuler le comportement d'une onde radar dans le sous-sol de Mars, mais aussi les interactions avec la structure du rover. Ce type de simulation numérique sera nécessaire pour calibrer et interpréter les futurs donnés de WISDOM.

L'objectif des travaux que nous présentons consiste à étudier des facteurs pouvant provoquer des perturbations des mesures, en vue de les corriger. Dans un premier temps nous présenterons la solution que nous avons retenu afin de pouvoir lors d'un même calcul simuler le déplacement du rover sur plusieurs dizaines de mètres. Nous présenterons aussi les résultats relatifs à l'effet de l'inclinaison de l'antenne lors du déplacement dû à la topographie du site d'observation.

## 2   Présentation et exploitation du modèle numérique

Le système radar embarqué sur le rover utilisera une antenne cornet complexe qui intègrera des antennes vivaldi. Cette antenne permettra d'effectuer des mesures sur une bande de fréquence allant de 500 MHz à 3GHz. Cette antenne aura une incidence de 10° avec le sol lorsque le rover se trouve sur un sol plat et se situera à environ 50 cm du sol martien.

Notre objectif est de mettre au point un modèle numérique permettant de simuler le déplacement du radar. Le choix de l'antenne n'est pas définitivement fixé. Dans un premier temps, nous utiliserons donc comme antenne un simple cornet dont la bande passante varie de 900 MHz à 2.7 GHz. Par la suite cette antenne sera remplacée par l'antenne réelle qui est en cours de définition et de réalisation à l'université de Dresde.

Le logiciel de simulation retenu pour notre application est le logiciel HFSS (High Frequency Structure Simulator) de la société ANSOFT. Il utilise la méthode des éléments finis et se prête tout à fait à notre étude car il fonctionne dans le domaine fréquentiel et il pourra représenter convenablement le fonctionnement du radar qui travaillera en mode « step-frequency ».  La simulation de l'antenne cornet et éventuellement du rover lui-même est tout à fait envisageable à l'aide d'un ordinateur de type PC possédant au moins 512Mo de RAM.

Le milieu géologique sera constitué dans un premier temps d'une couche d'un mètre de poussière ($\varepsilon_r$=6.86-0.32j) recouvrant une couche de basalte humide ($\varepsilon_r$=15.28-1.05j). Ces deux milieux ne sont pas magnétiques ($\mu$=1).

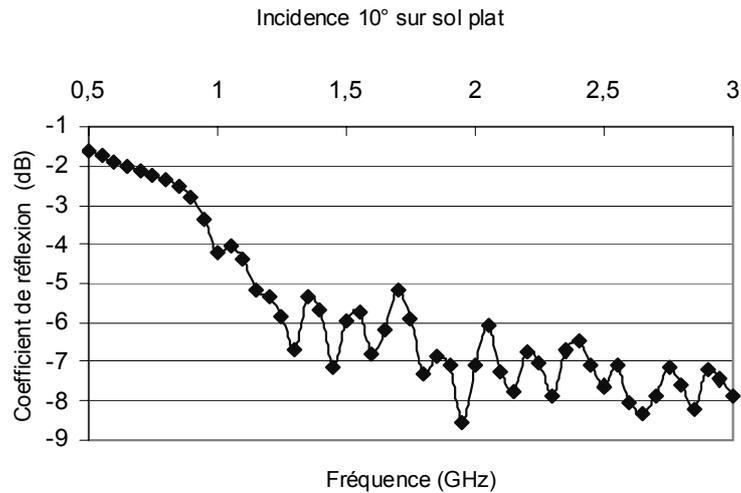

**Figure 1 : Coefficient de réflexion lors d'un déplacement sur un sol lisse et homogène.**

D'autres paramètres comme la topographie, la porosité de surface, la présence de gradients d'humidité, de roches … seront pris en compte par la suite. Les propriétés du milieu vont donc varier lors du déplacement du rover. Notre problème a donc consisté à simuler ce déplacement et donc celui de l'antenne au dessus d'une structure géologique dont les propriétés varient. Il n'est pas possible d'envisager une dichotomie du calcul par découpage de l'espace traversé par le rover : cela nécessiterait de relancer le calcul un nombre élevé de fois, ce qui gênerait considérablement l'exploitation des résultats. De plus, ces calculs devraient prendre aussi en compte la non uniformité du milieu observé par l'antenne lors d'une acquisition comme le montre la Figure 1.

La solution pour représenter la structure géologique complète pourrait consister à la représenter entièrement à l'aide de notre modèle numérique. Malheureusement la mémoire informatique nécessaire rend cela impossible. La solution que nous avons retenue consiste à créer un modèle représentant l'antenne du radar et la zone d'observation de l'antenne lors de la mesure. Le maillage utilisé pour représenter ce système ainsi que le temps de calcul nécessaire restent alors raisonnables. La prise en compte de la variation du milieu observé lors du déplacement s'effectue par variation des propriétés (taille, permittivité…) durant le calcul. HFSS possède une option de calcul paramétrique, le calcul étant alors effectué sur une gamme de variation de paramètres. La permittivité du milieu peut ainsi être fonction d'un paramètre correspondant au déplacement du véhicule. La zone observée est décomposée en différentes parties afin de tenir compte de la non homogénéité du milieu comme le montre la Figure 2.

Il nous est alors possible de faire varier l'épaisseur des couches, leur rugosité de surface ou encore la permittivité des milieux en fonction d'une variable représentant le déplacement du rover.

Nous présentons ci-dessous les résultats en changeant la permittivité du milieu en fonction du déplacement de l'antenne. Nous avons introduit dans Excel des valeurs de la permittivité pour différentes positions de l'antenne (« D » représente la distance de l'antenne à sa position originale). Nous avons obtenu une courbe de tendance pour $\varepsilon(D) = f(D) = \varepsilon'(D) + j\,\varepsilon''(D)$ :

$$\varepsilon' = -3 \cdot 10^{-5} \cdot D^4 + 0.0023 \cdot D^3 - 0.052 \cdot D^2 + 0.0706 + 10.009$$
$$\varepsilon'' = 9 \cdot 10^6 \cdot D^4 - 0.0006 \cdot D^3 + 0.0159 \cdot D^2 - 0.2249 \cdot D + 1.991$$

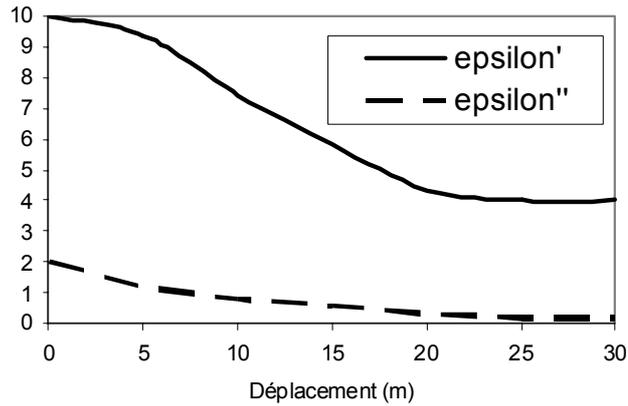

**Figure 2 : Variation de la permittivité lors du déplacement.**

Les calculs ont été effectués sur toute la bande de fréquence mais nous présentons ci-dessous le résultat du coefficient de réflexion S11 calculé au niveau de l'antenne pour une fréquence de 1.4 GHz (milieu de bande).

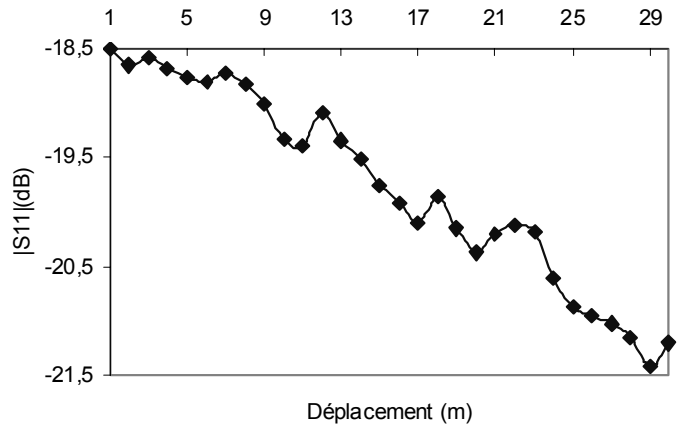

**Figure 3 : Coefficient de réflexion lors du déplacement sur un sol lisse non uniforme.**

Par endroit, le rover peut traverser des terrains où les sols sont très différents. Par exemple une partie du terrain peut comporter des inclusions (roches). L'apparition d'inclusions à divers endroits du trajet est prise en compte dans notre modèle à l'aide de fonctions de type « tout ou rien ». L'inclusion apparaît dans le modèle uniquement lorsque qu'elle apparaît dans la zone d'observation du rover lors du déplacement.

Le rover se déplacera au dessus de zones dont la topographie peut varier. Notre modèle peut introduire ces variations de topographie. Toutefois ce modèle a du être perfectionné en vue de prendre en compte l'inclinaison du rover lors du déplacement comme le montre la Figure 4.

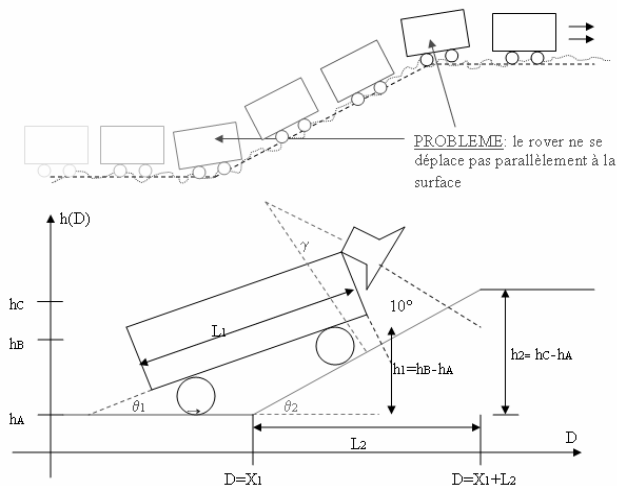

**Figure 4 : Prise en compte de la topographie.**

Nous avons introduit dans notre modèle la possibilité de fixer la position de chaque roue lors du déplacement du rover en fonction de la topographie rencontrée. Un algorithme permet de recalculer automatiquement le nouvel angle d'incidence de l'antenne et prend en compte l'inclinaison du rover. La Figure 5 présente les premiers résultats obtenus lors de l'introduction d'un petit dénivelé. Ce dernier est une bosse dont la hauteur varie et dont la base est large de 8 m. Cette bosse est placée à 6 m du point de départ du rover (D=0). Les résultats présentés sur la Figure 5 nous permettent d'analyser l'influence de l'inclinaison du rover sur les mesures. Si nous considérons qu'une perturbation supérieure à 2 dB ne doit pas être négligée alors nous pouvons constater que cette condition est respectée pour une inclinaison inférieure à 25% pour les situations A et C (voir Figure 5). Malheureusement cette condition ne peut être respectée que pour des inclinaisons de 13% pour les configurations de terrain C.

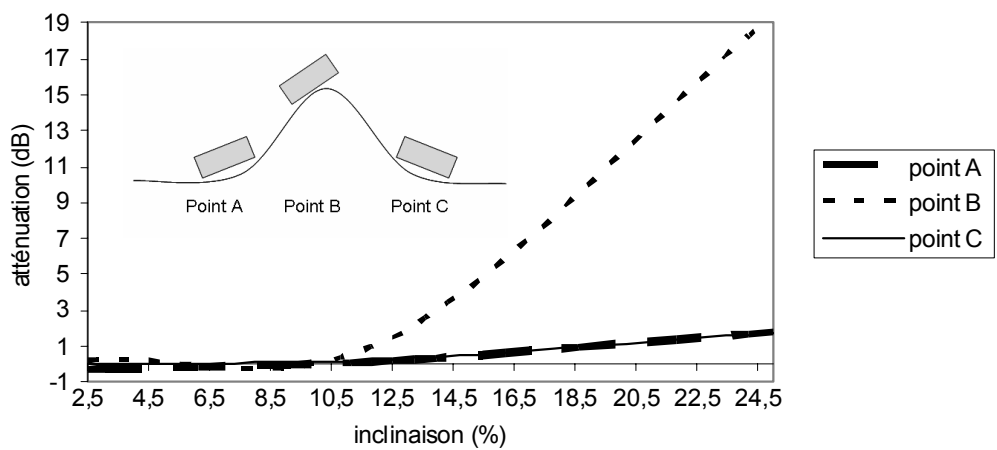

**Figure 5 : Atténuation provoquée par l'inclinaison du rover en fonction du pourcentage de la pente.**

### 3 Conclusion

Nous avons développé un modèle qui simule correctement le fonctionnement du radar lors de son déplacement. Les premiers résultats que nous avons obtenus nous ont permis de mettre en évidence la sensibilité des mesures à la variation de la permittivité des milieux et à la topographie de surface. Ce dernier point pourrait conduire à lancer une réflexion sur l'exploitation des futurs résultats de la mission. Une correction d'angle d'incidence ou une correction des mesures pourraient ainsi être envisagées. Sachant que des milieux très complexes pourront être rencontrés sur Mars, nous allons continuer à enrichir notre modèle en lui ajoutant la possibilité d'introduire de nouveaux paramètres tels que la porosité de surface ou des gradients de permittivité. Nous devrons aussi prendre ne compte l'antenne qui sera définitivement choisie ainsi que la structure du rover.